\documentclass{aa}
\usepackage{graphicx}
\usepackage{verbatim}
\usepackage{bigdelim}
\usepackage{textcomp}
\usepackage{color}
\usepackage{afterpage}
\usepackage{longtable}
\usepackage{enumerate}
\usepackage{multirow}
\usepackage{pdflscape}
\usepackage[square,authoryear]{natbib}
\usepackage[margin=0.7in]{geometry} 
\bibpunct{(}{)}{;}{a}{}{,} 
\usepackage{verbatim}
\usepackage{microtype}
\usepackage{fixltx2e} 
\usepackage{amsmath}
\usepackage{amssymb}

\title{A more detailed look at Galactic magnetic field models:\\ using free-free absorption in HII regions} \titlerunning{three dimensional synchrotron emissivity}
\author{I.M. Polderman \inst{\ref{inst1}}
\and M. Haverkorn \inst{\ref{inst1}}
\and T.R. Jaffe \inst{\ref{inst2}, \ref{inst3}}}

\authorrunning{Polderman I.M. et al.}
\institute{Department of Astrophysics/IMAPP, Radboud University, P.O. Box 9010,6500 GL Nijmegen, The Netherlands.\label{inst1}
\and 
CRESST II, NASA Goddard Space Flight Center, Greenbelt, MD, 20771, USA.\label{inst2}
\and
Department of Astronomy, University of Maryland, College Park, MD, 20742, USA.\label{inst3}
}

\abstract
{Cosmic rays (CRs) and the Galactic magnetic field (GMF) are fundamental actors in many processes in the Milky Way. The observed interaction product of these actors is the Galactic synchrotron emission integrated over the line-of-sight (LOS). A comparison to simulations can be made with this tracer using existing GMF models and CR density models. This probes the GMF strength and morphology and the CR density.}
{Our aim is to provide insight into the Galactic CR density and the distribution and morphology of the GMF strength by exploring and explaining the differences between simulations and observations of synchrotron intensity. }
{At low radio frequencies HII regions become opaque due to free-free absorption. Using these HII regions we can measure the synchrotron intensity over a part of the LOS through the Galaxy. The measured intensity per unit path length, i.e. the emissitivity, for HII regions at different distances, will allow probing variation in synchrotron emission not only across the sky but also in the third dimension of distance. Performing these measurements on a large scale is one of the new applications of the window opened by current low frequency arrays. Using a number of existing GMF models in conjunction with the Galactic CR modeling code GALPROP we can simulate these synchrotron emissivities.}
{We present an updated catalog of low-frequency absorption measurements of HII regions, their distances and electron temperatures, compiled from the literature. We report a simulated emissivity that shows a compatible trend for HII regions that are near to the observer. However, we observe a systematically increasing synchrotron emissivity for HII regions that are far from the observer, which is not compatible with the values simulated by the GMF models and GALPROP.}
{State-of-the-art GMF models plus a GALPROP generated CR density model cannot explain low-frequency absorption measurements. One possibility is that distances to all HII regions cataloged at the kinematic 'far' distance are erroneously determined, though this is unlikely since it ignores all evidence for far distances in the literature. However, a detection bias due to the nature of this tracer requires us to keep in mind that certain sources may be missed in an observation. The other possibilities are an enhanced emissivity in the outer Galaxy or a diminished emissivity in the inner Galaxy.}

\begin{document}
\maketitle
\section{Introduction }
\label{sec:intro}

Magnetic fields are prevalent in the Milky Way. They are present in Galactic sources such as supernova remnants, stars and pulsars \citep{Wielebinski2005}. But they are also present in the tenuous medium that is ubiquitous in the Galaxy, the interstellar medium (ISM). The ISM consists of gas and dust, cosmic rays (CRs) and magnetic fields. Even though it makes up only a small part of the total Galactic mass, it makes a vital contribution to different processes that take place in the Galaxy, and is essential to the Galactic `ecosystem' \citep{Ferriere2001}. Since it influences the evolution of galaxies, it is important to understand and quantify the properties of the ISM \citep{Heiles2012}. In this paper we will not discuss the gaseous part of the ISM, but focus on the Galactic magnetic field (GMF) and the CRs, whose energy densities comparable to that of the turbulent interstellar gas.

Because the magnetic fields are involved in a variety of processes they are of interest to a large number of fields in astronomy and astrophysics. In addition, a better understanding of the GMF will allow a better understanding of the polarized foregrounds that challenge the communities measuring the Cosmic Microwave Background and the Epoch of Reionization. Yet another community is interested in particles with the highest energies in the Universe, the ultra-high-energy cosmic rays, and needs to understand the deflection of these particles in the GMF before they are measured on Earth. This is important if they want to back trace these particles to their sources.

When the discovery of starlight polarization in the first half of the 20th century \citep{Hall1949,Hiltner1949b,Hiltner1949a} hinted at the existence of a magnetic field in the Milky Way, the field of Galactic magnetism was set in motion. Since then an enormous body of work has been created in an effort to understand it. The subject has been approached from theoretical, observational and numerical simulation points-of-view and here we build on work that spans decades. For recent reviews see \citet{Haverkorn2015} and \citet{Jaffe2019}. 

Because we are inside the Milky Way it is difficult to disentangle the different actors that produce observational tracers, and some clever tricks have to be implemented to extend our measurements (which are projected onto the sky) into a three dimensional view of our Galaxy. This unfortunately means that sometimes the GMF properties that are available in the models are physically unlikely. Most recently a different approach is being made by \citet{Shukurov2019} with which it is possible to calculate the magnetic fields from dynamo theory. This is a valuable addition to the existing GMF models, e.g.,
 \citep{Jansson2012a,Jansson2012b},
 \citep{Sun2010},
 \citep{Jaffe2013},
 \citep{Fauvet2011},
 \citep{VanEck2011}, \citep{TT2002} and \citep{Page2007}.
 
In this paper we would like to focus mostly on the observational approach. As a tracer we use the synchrotron emission that is the dominant radiation in the low radio frequency sky and is a product of the gyration of CREs around the magnetic field lines, and produces an intensity according to:

\begin{equation}
I \propto \int n_{CR}~B_{\bot}^{\frac{p+1}{2}} ~dL,
\label{eq:synchrotronintensity}
\end{equation}
where the exponent $p$ is assumed to have a value of three, as follows from the Cosmic Ray electron (CRE) spectrum \citet{Planck2016}. Clearly this tracer is a convolved effect of the interaction between the CRE density, $n_{\rm CR}$ and the GMF strength in the direction perpendicular to the line-of-sight (LOS),  $B_{\bot}$, integrated over the path length $dL$. To properly compare any observations to simulations we will need a model for both the CRE density and the GMF. However, we do not create our own CRE density or GMF models, we only want to compare the most current models to the catalog of data that we have available, see Sects. \ref{sec:theory} and \ref{sec:catalog}. In our previous paper \citet{Polderman2019} (hereafter P19), a set of GMF models was used in combination with a constant CRE density throughout the Milky Way. In this paper we have turned to the {\sc GALPROP} code \citep{Strong2009}. This code is able to calculate the propagation of relativistic charged particles and produce a Galactic CRE density model. With this combination of models we should be able to achieve a more realistic comparison to the observations.

In Sect.~\ref{sec:theory} we explain the theory behind the synchrotron tracer and its interpretation. Section~\ref{sec:catalog} discusses the updates that have been done to the catalog presented in P19 and the consequences for the emissivity distribution. In Sect.~\ref{sec:method} the method used is described and Sect.~\ref{sec:results} presents our results. The discussion of the results can be found in Sect.~\ref{sec:discussion}, and Sect.~\ref{sec:conclusions} contains our conclusions.

\section{Theory }
\label{sec:theory}

Observations of the Galactic plane at frequencies below 100~MHz open a window into a regime of Galactic synchrotron-dominated emission \citep[e.g,][]{Kassim1990,Odegard1986}. The emission we observe is the direct product of the interaction between CREs and the perpendicular component of the Galactic magnetic field, $B_{\bot}$, and can therefore be used to constrain either of these variables, or both, when making assumptions such as equipartition or pressure equilibrium. 

In this low-frequency regime the observed HII regions will be affected by free-free absorption, turning them into discrete absorption regions on the sky against the synchrotron background radiation. The brightness temperature measured in these absorbed HII regions is then a measure for the synchrotron radiation emitted in the LOS \textit{behind} this HII region. The following section explains this concept.

When performing observations of HII regions at low radio frequencies, and the measured brightness flux is absolutely calibrated, the following brightness temperature can be observed:

\begin{eqnarray}
T_{\rm obs} = T_{\rm F} +T_{\rm e} \,(1 - e^{-\tau})+T_{\rm B}\,e^{-\tau},
\label{eq:kassimfull}
\end{eqnarray}
consisting of: the foreground brightness temperature, $T_{\rm F}$, which is the brightness temperature of the LOS between the observer and the HII region; the electron temperature, $T_{\rm e}$, the temperature of the electrons in the HII region; the background brightness temperature, $T_{\rm B}$, the brightness temperature of the LOS between the HII region and the Galactic edge; and the opacity of the HII region, $\tau$. At low frequencies the high opacity approximation can be used, $\tau~>>~1$. This causes the equation to simplify to:

\begin{eqnarray}
T_{\rm obs} = T_{\rm e} + T_{\rm F}.
\label{eq:approxsd}
\end{eqnarray}

With a measured electron temperature the {\it foreground} brightness temperature can be inferred from the observation. In our catalog we have several of these HII regions. However, by far the most observations in our sample are performed interferometrically, and without an absolute flux calibration. The only quantity that can reliably be measured in this way is the deficit relative to an unknown smooth distribution. The equation for this can be calculated by taking eq.~\ref{eq:approxsd} and subtracting the large scale structure that is missed by the observer. This large scale, or total, brightness temperature is defined as: $T_{\rm T}=T_{\rm F}+T_{\rm B}$.

\begin{eqnarray}
T_{\rm obs,IF} &=& T_{\rm e} + T_{\rm F} - T_{\rm T}\\
&=& T_{\rm e} - T_{\rm B}
\label{eq:approxint}
\end{eqnarray}

This method provides the brightness temperature of the emission in the LOS behind the HII region from the observer perspective, in other words {\it background} measurements. These brightness temperatures still depend on the length of the line of sight (path length) over which they were integrated by the observation. The quantity in which we are interested is the emission per path length, and can be calculated for both the foreground and background brightness temperatures. Where $D_{\rm F}$ is the distance to the HII region, and $D_{\rm B}$ is the LOS distance from the HII region to the Galactic edge, throughout the paper we will use the term path length for this quantity. This path length needs to be calculated from the distance to the HII region using a distance from the Earth to the Galactic center of 8.5~kpc. We assume that the Galaxy has a cut-off at the Galactic radius of 20~kpc as is assumed in both the GMF models and the GALPROP code, the consequences of which for our dataset are discussed in Sect.~\ref{sec:notreal}.
We can then proceed to calculate the emissivities as,
\begin{eqnarray}
\epsilon_{\rm F}&=&T_{\rm F}/D_{\rm F}\\
\epsilon_{\rm B}&=&T_{\rm B}/D_{\rm B}.
\label{eq:emiss}
\end{eqnarray}

where $\epsilon_{\rm F}$ is the average emissivity on the LOS between the observer and the HII region, and $\epsilon_{\rm B}$ is the average emissivity along the LOS behind the HII region. In the rest of the paper the term `emissivity' will always indicate the average emissivity over a LOS.

A set of these emissivities can be seen as three dimensional data. It takes two coordinates to specify the HII region location in the plane of the Milky Way, while the third holds information on the integrated emissivity of the path length behind the HII region.

A strict requirement for the HII region is size. To meet this requirement the HII region has to be larger than the beam, otherwise the measurement will be contaminated by unwanted synchrotron emission. \citet{Su2018} used a different method to calculate the foreground and background emissivity for some of these sources, but in this paper we choose to follow \citet{Kassim1990}.


\section{Catalog}
\label{sec:catalog}
In P19 we presented a catalog of foreground and background emissivity measurements from five literature sources \cite[e.g.,][]{Jones1974,Roger1999,Nord2006,Hindson2016,Su2016}. It contained 9 foreground measurements and 115 background measurements. For proper comparison, all the catalog entries were rescaled to 74~MHz. The longitude range of the catalog is --43$^{\circ} < \ell < $ 25$^{\circ}$. The specific updates to the catalog are discussed in the section below and the changing distribution in the emissivities is discussed as well.

\subsection{Updates to the Polderman et al (2019) catalog}
\label{subsec:updates}

A significant part of the data in the P19 catalog comes from the work of \cite{Nord2006} (hereafter N06). For some of these sources, new distance information is now available. Therefore, we update both the amount of sources from N06 and their distances in our new catalog, in three ways.

First, source distances in N06 are obtained from the HII region catalog by \cite{Paladini2003}, who calculate kinematic distances using IAU standard values for the Sun's Galactocentric radius and velocity in the Milky Way. Of their 458 sources with a distance ambiguity, the distance ambiguity can be solved for 117 sources using auxiliary data, viz.\ absorption lines (HI, H$_2$CO or OH) or optical counterparts. This means that 281 HII regions in the Paladini catalog are left with a distance ambiguity. This ambiguity is analyzed with a luminosity-physical diameter correlation, which has such a high scatter that it does not resolve the issue for individual sources, but only gives statistical information for the whole sample.
  P19 includes N06 sources where the distance ambiguity was only resolved statistically. Here, we revisit those sources and include these if distance estimates in other catalogs were present (see below). For 9 sources without any resolution to the ambiguity, no newer distance estimates are available, so these sources are discarded.

Second, we compare source distances in P19 with distance estimates in 5 more recent HII region catalogs, i.e. \cite{Reid2014, Balser2015, Quireza2006, Anderson2014} and \cite{Anderson2018}. \cite{Reid2014} use parallax measurements for distance estimation, which we deem most reliable. The other catalogs use kinematic distances, in which the distance ambiguity is resolved through absorption lines and/or optical counterparts. For five N06 sources, parallax measurements are available, which are adopted as new distance measurements.
For three sources, updated kinematic distances are available which are all consistent with the original N06 distances; for these, we use the original N06 distances

Third, we include 30 additional N06 sources which did not have known distances in 2006, but have more recent distance estimates. For distance determinations that do not have any errors given, we assume an error of $50\%$.

Though our catalog contains foreground measurements, we will not use them in the rest of this work. Since we assumed a local emissivity enhancement in P19 to be able to explain the observations, and not all the GMF models have this, it would complicate the comparisons we present in this work. The total number of background measurements we use in this paper is 135.

\subsection{Updated emissivity distribution}

\begin{figure}

\resizebox{\hsize}{!}{\includegraphics{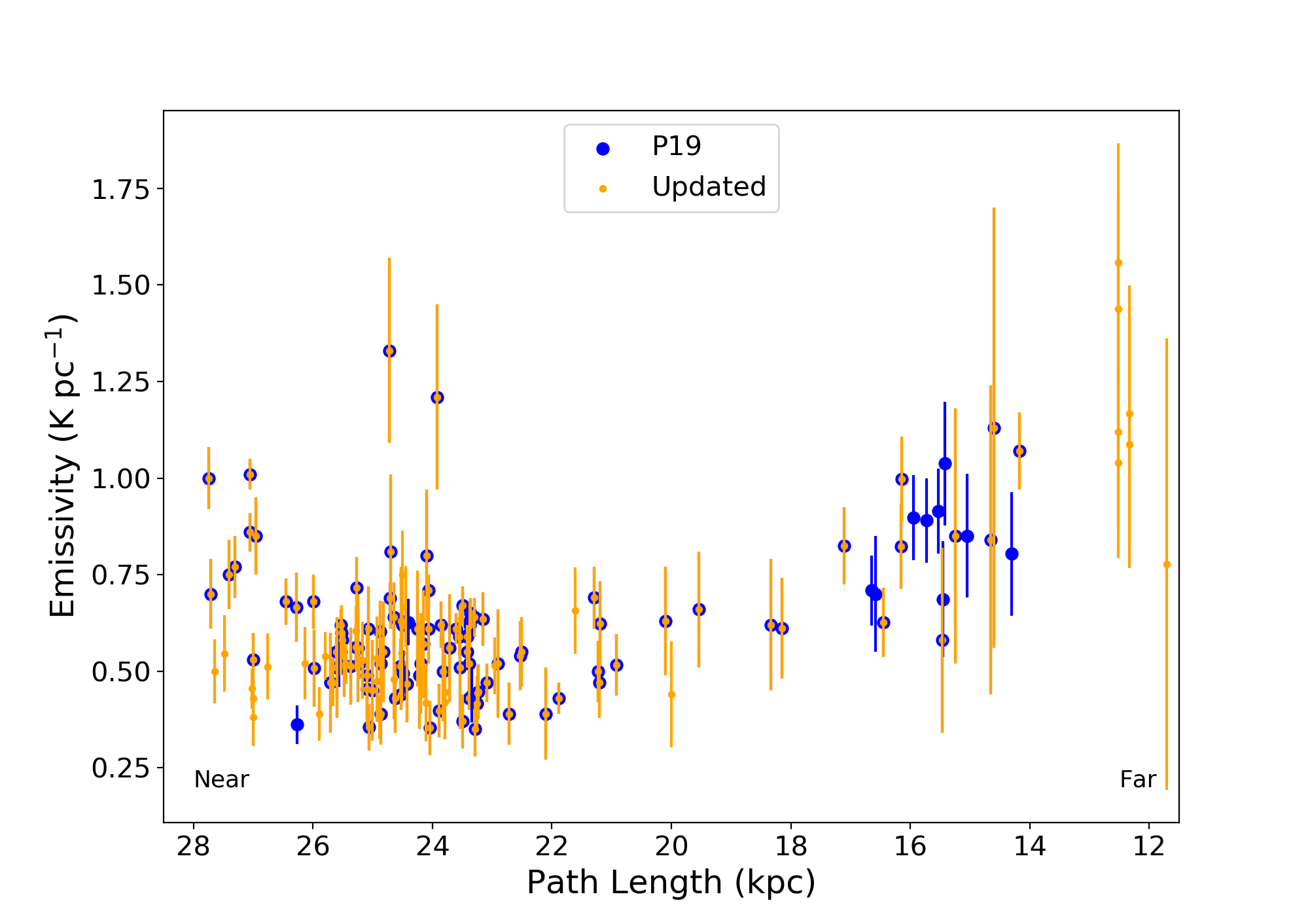}}
  \caption{ The background emissivity (K~pc$^{-1}$) as a function of path length (kpc) from the considered HII region to the far boundary of the Galaxy, with the P19 data in blue and the updated catalog data in smaller orange dots. The error bars include propagation of the brightness temperature error and the distance error.  NB, the x-axis is in reversed order with respect to path length, with short path lengths on the right and long path lengths on the left. }
  \label{fig:comparingCatalogs}
\end{figure}

There are several clear differences between the two datasets that can be seen in Fig.~\ref{fig:comparingCatalogs}. We go through the differences from left to right in the plot, which starts at the long path lengths in the bottom axis. The long path lengths correspond to HII regions that are near the observer.
Five HII regions with path lengths around 28~kpc are added in the new catalog, with emissivities around 0.50~K~pc$^{-1}$ that fit better with the emissivities for slightly shorter path lengths. 
Between 22 and 16~kpc the two added HII regions seem to fit within the trend that is set by the P19 HII regions. 

These HII regions bridge the area between the bulk of the `near' (28-22~kpc) and `far' (18-14~kpc) populations, where the `near' (`far') indicates the solution to the kinematic distance determination with the shortest (longest) distance from the Sun to the HII region. The `far' population has a higher mean emissivity than the `near' population. 
Around 16~kpc many of the HII regions from the P19 dataset are removed. The distances for these HII regions were ambiguous and were discarded.

The new data set has more HII regions that are farther away, at path lengths around 12~kpc. The emissivities for these HII regions are consistent with a rising trend in emissivities that is observed in the `far' population in the previous catalog as well. For the HII regions around 12~kpc the error margins are larger than for the other HII regions, this is due to having to assume an error of 50\% in the absence of an error in the literature.

\section{Method}
\label{sec:method}
In this section we discuss the different steps undertaken to simulate emissivities, and the comparison to the catalog. We use the {\sc Hammurabi} code \citep{Waelkens2009} to simulate the emissivity in the direction of the different HII regions. For this effort we use three GMF models that we discuss briefly in sect.~\ref{subsec:GMF}, and in addition to that we use the {\sc GALPROP}\footnotemark[1]\footnotetext[1]{http://galprop.stanford.edu} code \citep{Strong2009} to simulate the CRE density in the Milky Way, as discussed in Sect.~\ref{subsec:CREmodels}.  

\begin{table*}
\caption{Galactic magnetic field models}              
\label{tab:GMF}      
\centering                                      
\begin{tabular}{*{6}{l}}          
\hline\hline                        
Model & coherent & ordered & isotropic& halo & other \\
&&random&random&&\\
\hline 
\hline
    J13b & 4 spiral arms + &scaled to&\checkmark&\checkmark&molecular ring\\
    & inter-arm regions  &isotropic random&&``thick" disk&\\
    &&component&&&\\
    \hline
    JF12b & 8 spiral segments &scaled to &\checkmark&\checkmark&central 5~kpc \\
    &&coherent &&x-shaped&hole\\
    &&component&&&\\
    \hline
    Sun10b & field strength changes &no &\checkmark &\checkmark&Galactocentric ring\\
    & in concentric circles,&&&``thick" disk& around inner 5~kpc \\
    &field direction has pitch angle&&&&\\

\hline
\hline 
\end{tabular}
\end{table*}

\subsection{Galactic Magnetic Field models}
\label{subsec:GMF}

In this work we simulate emissivities by using three different GMF models, and compare these to the observed emissivities in the catalog. Table~\ref{tab:GMF} contains the general components of the GMF models. The GMF models we use are presented in \citet{Jaffe2016} as the JF12b, Sun10b and J13b models. In \citet{Jaffe2016} the parameters for these GMF models are adjusted to match the Planck data using a common CRE model. These models are based on the three models in \citet{Jansson2012a,Jansson2012b}, \citet{Sun2010} and \citet{Jaffe2013} respectively. More detailed information on the GMF model parameters and their determination can be found in \citet{Jaffe2016}. 

The parameter values for the models have been downloaded from the {\sc Hammurabi} sourceforge page\footnotemark[2]\footnotetext[2]{https://sourceforge.net/projects/hammurabicode/files/\\supplementary/}. 
Some changes were made to the parameter files due to the nature of the set-up used in \citet{Jaffe2016}, where the nearest 2~kpc to the Sun was simulated with higher resolution than the rest of it. More specifically we have made use of the ``RG2" parameter files and have adapted them such that all parts of the plane (min\_Radius=0) were integrated.

\subsection{Cosmic ray electron models}
\label{subsec:CREmodels}
The {\sc GALPROP} code \citep{Strong2009} is developed for the propagation of relativistic charged particles and to simulate diffuse emission that is produced during that propagation. In this work we do not make use of the latter function, but only use {\sc GALPROP} to calculate a CR density in the Milky Way. This happens in a self-consistent way using the magnetic field model that is employed to calculate the resulting synchrotron emissivity. The {\sc GALPROP} code uses a configuration file (GALDEF), that specifies source distribution and boundary conditions for the different CR species, to solve the transport equation. The code includes processes like convection/Galactic wind, diffusive reacceleration, energy loss and radioactive decay, among others. It is important to note that the diffusion is treated isotropically in this code. For consistency with \citet{Jaffe2016} we have used the GALDEF `z10LMPDE', based on \citet{Orlando2013}. This specific model is the result of a significant body of work of the references therein to develop a model for the spatial and spectral distribution of cosmic ray leptons. It has updated scale heights for the leptons and updated magnetic field parameters, and its spectrum has been specifically adjusted to fit the Fermi direct measurements of electrons and positrons, as well as the diffuse gamma-ray emission. A further improvement sees a better fit to the distribution of synchrotron emission in Galactic longitude and latitude. Further details can be found in the papers mentioned above.

For completion we also discuss the results when using a different GALDEF, to wit, the `71xvarh7S' model \citep{Abdo2010}, which is based on a different distribution of CR sources, particularly in the outer Galaxy. 
The GALDEF was run both in two-dimensional and three-dimensional mode. No significant difference between the runs was detected for any model. The three-dimensional runs are presented in this paper. 

\subsection{Modeling the emissivities behind the HII regions}
We use {\sc Hammurabi} \citep{Waelkens2009} to simulate the synchrotron intensities along specific lines of sight in the Milky Way. To simulate this tracer we need to provide {\sc Hammurabi} with a CR density and GMF models as discussed above.

For each HII region the synchrotron intensity is calculated for two parts of the LOS, one intensity integrated over the path from the observer to the HII region (T$_{\rm F}$), and the other integrated over the path between the Sun and the edge of the Milky Way (T$_{\rm T}$). Subtracting the former from the latter a value for the intensity integrated over the path between the HII region and the Milky way edge is determined, this is the background simulated measurement. With simple trigonometric calculations, and the assumption of a distance from the Sun to the Galactic center of 8.5~kpc and a Galactic radius of 20~kpc, the length of the path between the HII region and the edge of the Milky way can be calculated. Dividing the synchrotron intensity by its companion path length, the synchrotron emissivity for a LOS is computed.

\section{Results}
\label{sec:results}
In this section we discuss the distribution of the catalog emissivities compared to the simulated emissivities for three different GMF models and two different CRE models.

\begin{figure}
  \resizebox{\hsize}{!}{\includegraphics{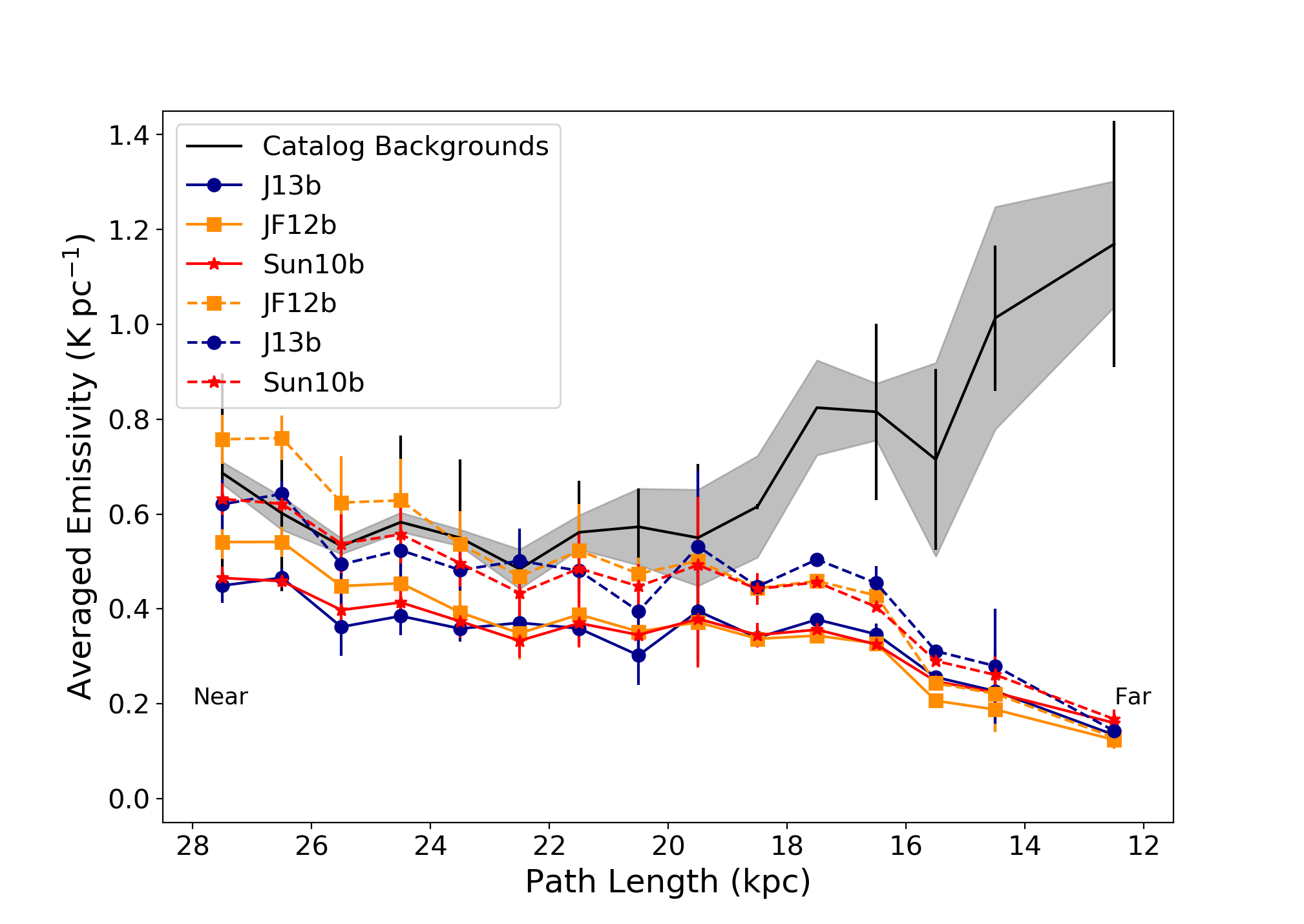}}
    \caption{ The averaged background emissivity (K~pc$^{-1}$) as a function of the path length (kpc) from the considered HII region to the far boundary of the Galaxy. The catalog data is shown in black with error margins in grey. This margin consists of the propagated measurement error for the data points in the bin. Also shown is the standard deviation, $\sigma$, in the observations as black error bars. The simulated values for two different CRE models are shown for the three different GMF models J13b, JF12b and Sun10b. The solid line indicates the results for the `z10LMPDE' CRE model and the dashed line indicates the results for the `71Xvarh7S' CRE model. All data - both simulated and catalog values - are averaged into bins with 1~kpc width. The standard deviation for the GMF models is calculated per bin and is plotted.}
\label{fig:ObsGMF}
\end{figure}

Fig.~\ref{fig:ObsGMF} shows the averaged emissivity as simulated with the different GMF models, plotted without any kind of extra normalization. In P19 we applied an arbitrary scaling to these emissivities to match them with the data, due to an oversimplified cosmic ray distribution model. However, when using {\sc GALPROP} to simulate a realistic CR density such a normalization is not needed.

\subsection{Emissivities from observational catalog}

Discussing first the catalog data, the trend of the data is clearly shown. At the longer path lengths of 28 kpc the data points start with a downward trend, from emissivities of 0.7~K~pc$^{-1}$ down to 0.5~K~pc$^{-1}$ at 26~kpc. This suggests an enhanced contribution of emissivity to HII regions with path lengths between 28 and 26~kpc.

The distribution between 26 and 22~kpc seems to rise and fall again, however we infer that this is due to the sporadic sampling of HII regions for these bins and the fact that HII regions at different longitudes can have the same path lengths. The distribution in this path length region is consistent with a flat one. 

At path lengths below 20~kpc an upward trend starts, and around path lengths of 12~kpc a peak of 1.2~K~pc$^{-1}$ is reached. This confirms the results of P19, where we concluded a diminished emissivity in the region around the Galactic center must exist, and/or a high emissivity contribution to the HII regions  at the far side of the Galaxy at distances larger than a few kpc from the Galactic center.

The error margin for the HII regions at the short path lengths is relatively large.The assumption of a 50\% error for the 'far' HII regions (those that were without a given error in literature) influences this. On top of this we notice an increase in the relative error for larger distances, we surmise this is due to the growing uncertainty of the distance determination and the error on the brightness temperature with growing distance to the HII region.

\subsection{CRE models}

For the `z10LMPDE' CRE model (solid lines) we see clearly that they favor a lower emissivity at shorter path lengths. For the most part, the models show the same large-scale trend. The longer path lengths down to approximately 18~kpc show a flat trend, whereafter the models show a downturn that signifies the lower emissivity contribution in the region behind the Galactic center. This is explained by both a lower GMF strength from the models and a lower CR density from {\sc GALPROP} in this region of the Galaxy.

For the `71Xvarh7S' CRE model (dashed lines) we see similarly clearly that simulated emissivities at longer path lengths are comparable to the catalog emissivities. The JF12b model again shows a deviation from the other models for the longest path lengths, but moving to shorter path lengths it rejoins the other models around 24~kpc. Thereafter the models show a downturn similar to that of the solid lines to finish at a path length of 12~kpc with an emissivity that is too low to be comparable to the catalog emissivities. Again this is due to the lower GMF strength and the CRE density in this region. We include this older model (superseded by `z10LMPDE') simply to show that this discrepancy remains for all GMF models even when a different CR source and propagation model is used.  We will explore the impact of a larger variety of viable CR models in future work.

\begin{figure*}

  \resizebox{\hsize}{!}{\includegraphics{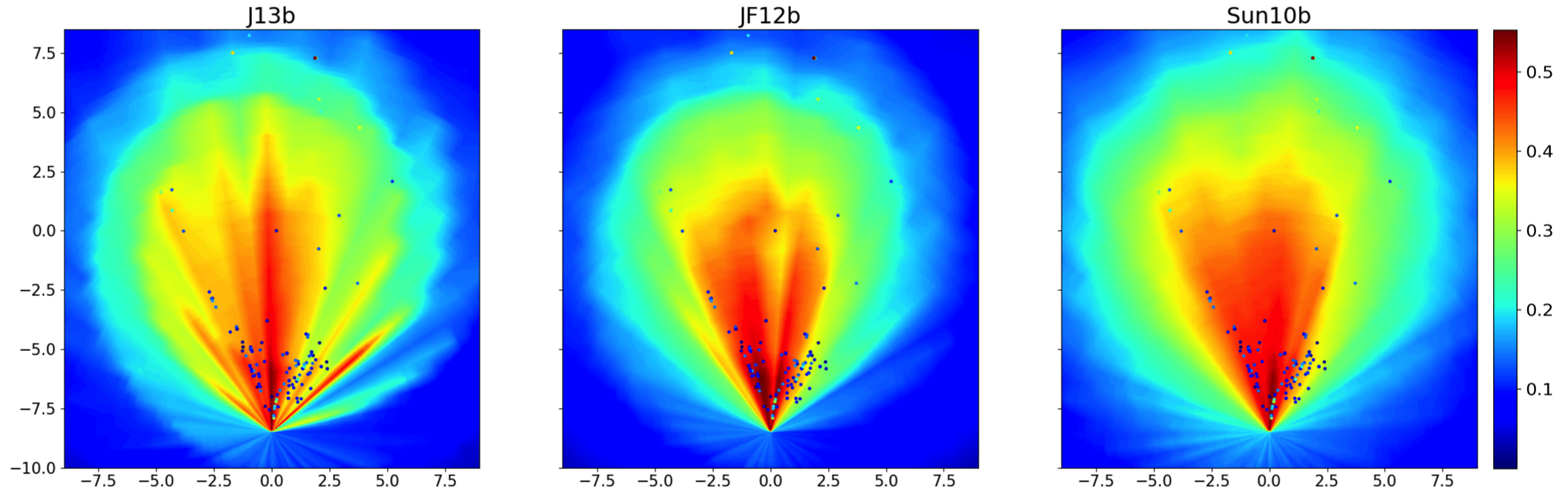}}
  \caption{Simulated background emissivities (K~pc$^{-1}$) in color scale with accompanying color bar, calculated for an evenly spaced grid of HII regions in the Galactic plane. The axes are in kpc. The Sun is located at (0,-8.5). Overplotted on all models are the observed background emissivities as colored circles. From left to right are plotted the three different GMF models.}
  \label{fig:GMFMock}
\end{figure*}

Different realizations of the turbulent magnetic field allow for minor variations in the emissivity that is simulated. We have investigated these minor variations by running several different realizations and calculating the difference in the output caused by each, and we find an estimated Galactic variance of 3 $\times 10^{-4}$~K~pc$^{-1}$. Such a small influence of the turbulent field is expected since it is on scales much smaller than what we probe. These fluctuations along the different LOSs are averaged out. A more extensive discussion on the Galactic variance can be found in \citet{Jaffe2016}.

In Fig.~\ref{fig:GMFMock} background emissivities spanning the Galactic plane are plotted for the three different GMF models and using the `z10LMPDE' CRE model. Only one realization of the turbulent magnetic field is shown here. In this view the Sun is located at xy-coordinates (0,-8.5). As was already seen in Fig.~\ref{fig:ObsGMF}, the emissivities simulated with the GMF models show a rapid decline in emissivity values with increasing Galactic radii. The structure of the GMF models is clearly visible, with the spiral arms apparent in all three figures and the slight asymmetry as a consequence of the pitch angle of the spiral arms.

\subsection{Comparison}
Comparing the observational and simulation results to each other we want to consider several points.
Firstly, it shows us that the simulated emissivity values for the z10LMPDE CRE model are only a factor of ~2 different in the mean from the observed values - though at the longer path lengths the difference is slightly smaller and at short path lengths the difference is slightly larger than this factor 2. And for the 71Xvarh7S the emissivities at the long path lengths are comparable to the catalog values within the error. This is both surprising and reassuring considering the uncertainty in the accuracy of both the CRE density and the GMF models. A possible explanation for the factor two could be found in the inaccuracy of the CRE spectrum at the energies that cause the low-frequency synchrotron emission due to the solar modulation. In Fig.~2 of \citet{Orlando2018} specifically it can be seen that at 74~MHz the models are not constrained, and could potentially cause the discrepancy.
Secondly, the GMF models themselves show results that are comparable to each other. We consider this reasonable due to the GMF parameter fit done in \citet{Jaffe2016}. 
Thirdly, the trends in the observations and the simulations are partially compatible. In the path length range from 28 to 18~kpc the relative flat distribution is reasonably well matched, but beyond 18~kpc the observational trend rises steeply upward, leaving the simulated trend behind.

\section{Discussion}
\label{sec:discussion}

In this section we will discuss different explanations for two general scenarios concerning the upturn in observational emissivity at short path lengths:
\begin{enumerate}
\item The upturn is \textbf{real}.
\item The upturn is \textbf{not real}.
\end{enumerate}
If the upturn is real the models are missing an essential piece of information. If it is not real it is an artifact of wrongly determined HII region emissivities, which can have several sources. Below we will discuss several arguments that favor either scenario.

\subsection{The upturn is real}
If the upward trend at short path lengths is real, it shows that the models underestimate the synchrotron emissivity in the far outer region and/or overestimate the emissivity in the inner region of the Milky Way. This does not subvert the current models nor the tracers used to constrain them. The discrepancy with current GMF models can simply be explained by the fact that this tracer probes different regions of the Milky Way than other tracers, i.e. the far Galaxy behind the Galactic center. Therefore it indicates the need for an extra component in either the CRE density or the GMF models: a component that results in an enhancement of emissivity in the outer region, or a paucity of emissivity in the inner region. 

If we consider any change in the emissivity in one region, we have to reflect on the effect it will have on the measurements. Any change affects the background measurements for HII regions whose LOS intersects with it. Nevertheless the change in emissivity does not translate into the same change for all such HII regions. Any enhancement in the outer Galaxy will change the emissivities for HII regions with shorter path lengths relatively more than for regions with longer path lengths. This is explained by the higher fraction of their path length that runs through this enhanced region. In this solution it is plausible that the emissivity keeps increasing with decreasing path length, if the \emph{fraction} of enhanced emission keeps increasing with decreasing path length. This could allow for the determination of a lower limit for the distance to this enhanced emission because it has to start behind the HII region that is the farthest away, otherwise the upward trend would have stopped. For our observational data this means the enhanced region would have to start beyond a distance of 16 kpc. 
This statement is only valid for the longitudinal range that is probed by our catalog. Any broader statements encompassing all longitudes will have to wait for future expansion of the catalog.

An enhanced emissivity could be the result of a higher CRE density or higher magnetic field strength (see Eq.~\ref{eq:synchrotronintensity}), but could also be the result of a more intermittent magnetic field.  As emissivity depends on B$_{\bot}^2$, a more clumped magnetic field will emit more synchrotron emission than a uniform magnetic field of the same average strength. Therefore, our data might be explained by an increasingly intermittent magnetic field in the outer Galaxy. Intermittent magnetic fields are indeed expected \citep{Seta2017}. However, the intermittency would have to be stronger in the outer Galaxy than in the inner region. We are not aware of any observational or theoretical evidence for this.
In addition, enhanced emissivity may also occur if the cosmic ray electron density and magnetic field are positively correlated, as discussed in \citet{Beck2003}. For this correlation to have any effect on the observations it has to be stronger in the outer regions of the Galaxy, like the intermittency. For this alternative, we are not aware of observational or theoretical evidence either.

Because we use background emissivity measurements any paucity in the inner region will not affect emissivities of HII regions that are beyond this inner region. In P19 we briefly discuss this region of diminished emissivity in the Galactic center region and put forward that it may be due to outflows and x-shaped magnetic fields in this area. Outflows are also described in \citet{Carretti2013}, and even if the size of the region they consider is an order of magnitude smaller than ours we still consider this option a possibility.

\subsubsection{Toy models}

\begin{figure}

  \resizebox{\hsize}{!}{\includegraphics{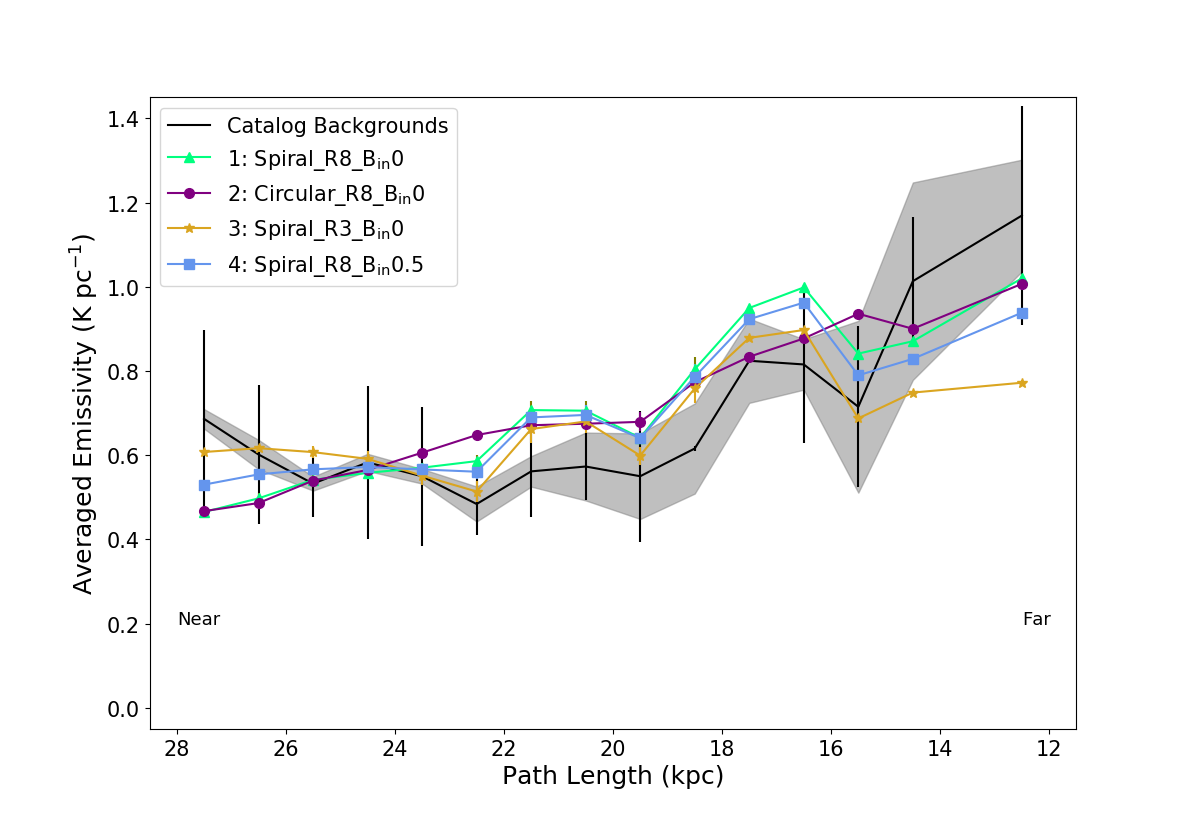}}
  \caption{ The background emissivity (K~pc$^{-1}$) as a function of path length (kpc) from the considered HII region to the far boundary of the Galaxy, with the catalog data in black with its error margin in grey and the four different toy models with model 1 in green triangles, model 2 in purple dots, model 3 in yellow stars and model 4 in blue squares.}
  \label{fig:toymodels}
\end{figure}

\begin{figure*}

  \resizebox{\hsize}{!}{\includegraphics{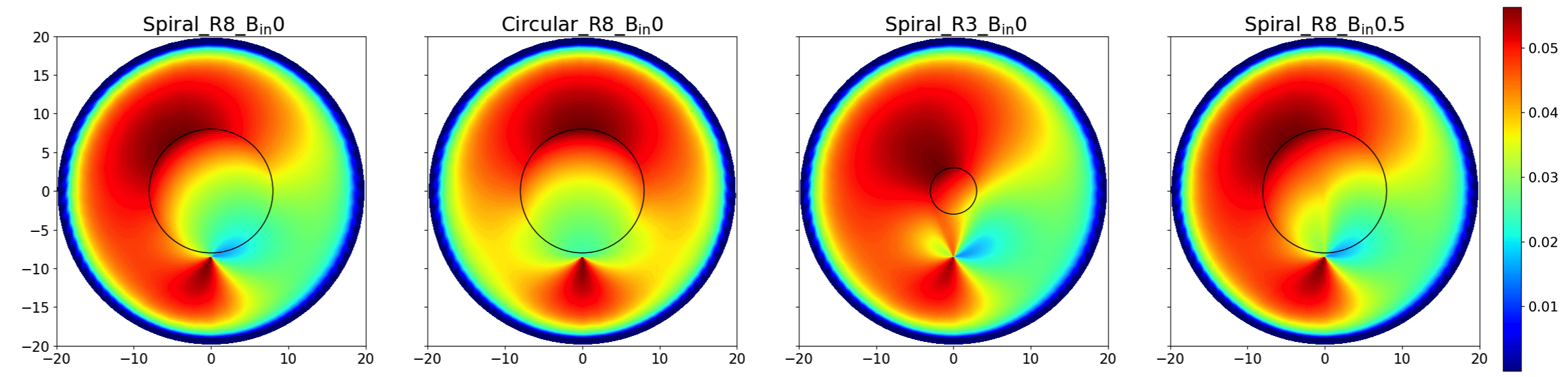}}
  \caption{Simulated background emissivities calculated for an evenly spaced grid of HII regions in the Galactic plane, the axes are in kpc. The Sun is located at (0,-8.5). From left to right: toy model 1, toy model 2, toy model 3 and toy model 4. For clarity we have plotted the radii for the inner region in black. No normalization of the values has been performed.}
  \label{fig:ToyMock}
\end{figure*}

\begin{table}
\caption{Toy models with an underdensity in the inner Galaxy. Columns (1) and (2) have the designation of each model, column (3) shows the radius of the inner region, column (4) has the magnetic field strength in the inner region, column (5) shows the magnetic field strength in the rest of the Galaxy and column (6) has the pitch angle for each model.}              
\label{tab:Toy}      
\centering                                      
\begin{tabular}{*{6}{l}}          
\hline\hline                        
\#&Model & R$_{\rm in}$ & B$_{\rm in}$ & B$_{\rm ou}$ & Pitch angle \\
&&kpc&$\mu$G&$\mu$G&degrees\\
\hline 
\hline
  1&  Spiral\_R8\_B$_{\rm in}$0 & 8.0 &0.0&1.0&-12.0\\
    \hline
   2& Circular\_R8\_B$_{\rm in}$0 &8.0 & 0.0&1.0&0.0 \\
    \hline
   3& Spiral\_R3\_B$_{\rm in}$0 & 3.0&0.0 &1.0 &-12.0\\
    \hline
    4&Spiral\_R8\_B$_{\rm in}$0.5 & 8.0&0.5&1.0&-12.0 \\
\hline
\hline 
\end{tabular}
\end{table}

In an effort to substantiate our hypotheses of an outer region with an enhanced emissivity contribution and an inner region with a diminished emissivity contribution we have created four simple toy models that exemplify this. To implement a toy emissivity model, we hold the CRE component constant over the Milky Way and vary the GMF strength for three models and the pitch angle for one. For the purposes of these simple toy models varying the GMF strength alone has the same effect on the synchrotron emission as varying the CRE density. All the details of the different models can be found together in Table~\ref{tab:Toy}. For all models, the outer 1 kpc is devoid of emissivity. This was chosen for numerical reasons near the boundary.

It is important to mention that models 1, 3 and 4 have a spiral structure due to the non-zero pitch angle. Model 2 however has a circular orientation of the magnetic field lines due to its pitch angle of zero degrees. The inner region for all the models is a circular region with a radius of 8~kpc for models 1, 2 and 4, and a radius of 3~kpc for model 3. Models 1, 2 and 3 have an inner circular region devoid of magnetic field strength, whereas model 4 has half that of the outer region.

Due to the arbitrary values of the magnetic field and the constant CRE density model, these toy models need to be normalized to the catalog data before we can start the comparison. We do this by employing a least-squares fit of the models to the catalog data. The result is shown in Fig~\ref{fig:toymodels}. 

Due to the normalization, the relative ratios of the models disappear. Therefore, we will confine ourselves to general statements about trends and features.

The most remarkable feature of the toy models is the upward trend, which is not an artifact of the normalization. Even though the constant CRE density and the limited magnetic field components used combine into a physically unlikely scenario, the contribution it provides to the shortest path length shows a striking resemblance to the observational trend. This leads us to our most important conclusion, that \emph{either a diminished emissivity in the inner Galactic region or an enhanced emissivity in the outer region} - or likely a \emph{combination of both} - can emulate the steepness of the observational trend at the shortening path lengths.

We can further discuss the different models by steepness of trend. Models 1 and 2 show the steepest of the trends, and the emissivity picked up for the longer path lengths is lowest. This makes sense because these path lengths, at least partially, run through the ``empty" region in the center and no emissivity is contributed in this part of the path length. The shorter path lengths are influenced least by the central region, which is a general remark for all four models. Model 4 is less steep than models 1 and 2. Its central region does have an emissivity contribution, so the longer path lengths pick up more emissivity which contributes to a flatter distribution. With the smallest radius of the inner region, model 3 has the flattest distribution. Here relatively more emissivity is picked up on the longer path lengths. We can conclude that the \emph{size of the inner region is correlated with the flattening of the distribution, and a smaller inner region (model 3) will likely not show the needed steepness to explain the trend in the observations. }

Finally, the influence of a pitch angle is clear: for model 2 (pitch angle of zero degrees) the emissivity distribution is much smoother and does not show the dips at path lengths of 15 and 22~kpc. This feature is also seen in the observational distribution. We conclude that for our uneven sampling \emph{a non-zero pitch angle is a prerequisite for any GMF model.}

The impact of the pitch angle is much clearer in Fig.~\ref{fig:ToyMock}. This can be seen in the results for model 2, which shows a symmetric emissivity distribution, where the rest of these models show a distinct asymmetry. It is this asymmetry that is unevenly sampled by our HII catalog that leads to the peaks and troughs in the curves in Fig.~\ref{fig:toymodels}

The toy models show other clear and understandable internal differences. Model 1 in Fig.~\ref{fig:ToyMock} shows strong emissivities in roughly the anti-center direction and in the first quarter beyond the inner region of 8~kpc that has no emissivity contribution. In this inner region the emissivity declines for HII regions that are closer to the observer, due to the increasing ratio of "empty" space over the path length. Model 3 has a smaller inner region, as a consequence of this we see in Fig.~\ref{fig:ToyMock} that more background emissivity is picked up for the HII regions in this center area. This is explained by a larger region where synchrotron emission can be produced.
Model 4 has an increased magnetic field strength in the inner 8~kpc, which can be clearly seen by the increased emissivity in this region. However, in our toy models a higher inner magnetic field strength causes less increase in the emissivity in this region than does a model that has a smaller inner empty region. 

\begin{figure}

  \resizebox{\hsize}{!}{\includegraphics{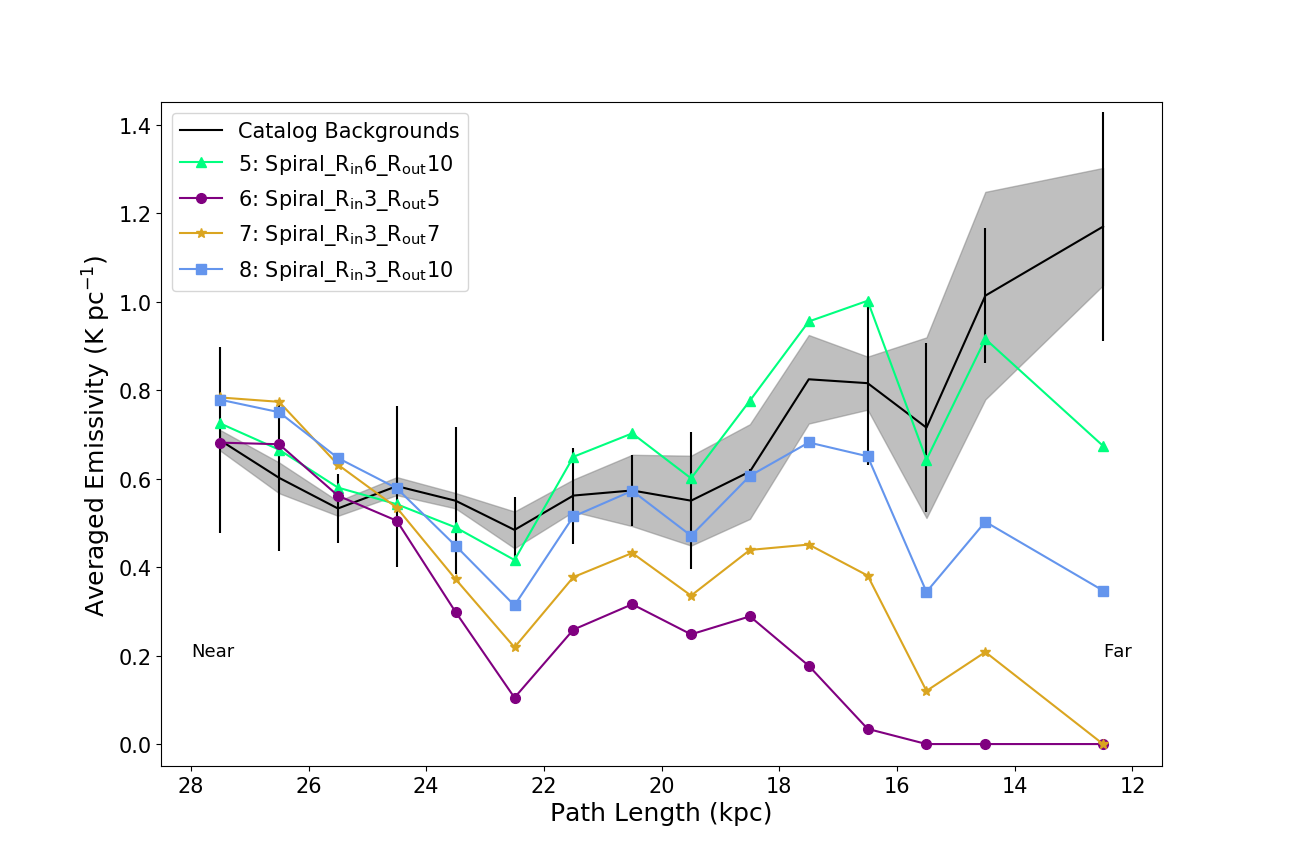}}
  \caption{ The background emissivity (K~pc$^{-1}$) as a function of path length (kpc) from the considered HII region to the far boundary of the Galaxy, with the P19 data in black with its error margin in grey and the four different toy models with model 5 in green triangles, model 6 in purple dots, model 7 in yellow stars and model 8 in blue squares.}
  \label{fig:toymodelsannulus}
\end{figure}

To constrain properties of the enhanced emissivity in the outer region of the Milky Way we use an additional set of toy models. The parameters for these models are described in Table~\ref{tab:ToyAnnulus} and the results are plotted in Fig.~\ref{fig:toymodelsannulus}. These models are annuli with different inner and outer radii, and are all centered on the Galactic center. Model 5 has an inner radius of 6~kpc and an outer radius of 10~kpc. Models 6 through 8 all have an inner radius of 3~kpc and increasing outer radii of 5, 7 and 10~kpc. For all these models the area that is covered by the annulus has a magnetic field strength of 1.0~$\mu$G and the area that is not covered by the annulus a magnetic field strength of zero. All models have a pitch angle of -12.0 degrees. Like for the other set of toy models, the CRE component is kept constant over the entire Milky Way, and a normalization of these toy models was performed. 

Depending on the width and location of the annulus many or few HII regions pick up emissivity. It is clear at the right hand side of Fig.~\ref{fig:toymodelsannulus} that the absence of emissivity in the outer region of the Galaxy impacts the HII regions with short path lengths. All the models show a decline or even absence of emissivity there. This cannot be reconciled with the upward trend of the observations.  Also clear is the impact of the size of the annulus, specifically for models 6 -- 8, which show an increasingly flattening trend with increasing annulus size. However the absence of emissivity behind the farthest HII regions will never allow these models to display a trend that is comparable to the observations. From this we are forced to conclude that \emph{any model used to describe the observations has a need for emissivity beyond a 10~kpc Galactic radius, in the far outer Galaxy.}

\begin{table}
\caption{Toy models with an underdensity in the inner Galaxy. Columns (1) and (2) have the designation of each model, column (3) shows the inner radius of the annulus, column (4) has the outer radius of the annulus, column (5) shows the values for the other parameters that do not vary in these models. These parameters are the pitch angle, the magnetic field strength in the annulus, B$_{\rm ann}$, and the magnetic field strength in the rest of the Galaxy, B$_{\rm nann}$.}              
\label{tab:ToyAnnulus}      
\centering                                      
\begin{tabular}{l l l l | l}          
\hline\hline                        
\#&Model & R$_{\rm in}$ & R$_{\rm ou}$& Other parameters  \\
&&kpc&kpc&\\
\hline 
\hline
  5&Spiral\_R$_{\rm in}$6\_R$_{\rm out}$10 &6.0& 10.0 &\\
    \cline{1-4}
6& Spiral\_R$_{\rm in}$3\_R$_{\rm out}$5 &3.0&5.0 & Pitch angle: -12.0$^{\circ}$ \\
     \cline{1-4}
 7&Spiral\_R$_{\rm in}$3\_R$_{\rm out}$7&3.0& 7.0&B$_{\rm ann}$: 1.0~$\mu$G  \\
     \cline{1-4}
8&Spiral\_R$_{\rm in}$3\_R$_{\rm out}$10&3.0 & 10.0&B$_{\rm nann}$: 0.0~$\mu$G \\
\hline
\hline 
\end{tabular}
\end{table}

\subsection{The upturn is not real}
\label{sec:notreal}

If the upward trend at short path lengths is not real, then either the radius of the Milky Way is wrongly chosen or there is a problem with the distance determination for HII regions. This section will discuss these options in detail.

To calculate the path lengths for the background emissivities a radius for the Galaxy has to be assumed. In previous work, and this work we have assumed that this radius is 20~kpc. Assuming a larger radius will adjust the emissivity values in the catalog in a downward fashion, though the trend remains the same. A smaller radius will drive the emissivity values up and will also not change the trend, additionally it will unnecessarily confine the region within which synchrotron emission could be produced. Considering the incompatible trends of the observations and the models, we looked into changing this variable to an arbitrary larger value of 30~kpc. We found that this procedure does not change the trend of our observations. 

The determination of kinematic distances of HII regions has the inherent issue of ambiguity: the region is either located at the near or the far distance solution. As discussed in Sect.~\ref{subsec:updates}, observations of other tracers can break this degeneracy. 
However, if we assume that this method does not always work, there might be HII regions that are currently categorized in the far distance, that are actually at the near distance. In our sample, a by-eye investigation reveals that roughly 9 HII regions with path lengths smaller than 20~kpc will have to move to counteract the upward trend. This would diminish the support for the rising trend on the right-hand-side of Fig.~\ref{fig:comparingCatalogs} and would bring the observed trend closer to the models. Taking this line of reasoning one step further, all but two HII regions with path lengths shorter than 20~kpc will have to move to make the catalog data compatible with the simulated emissivities in Fig.~\ref{fig:ObsGMF}. Unfortunately, this will leave most of the short path length regime unprobed, and no informative comparison can be made between the catalog and simulated emissivities. In addition, for some of these HII regions the distances were determined accurately with a different method, therefore this upward trend will not disappear.

Although it seems unlikely that far distances to all HII regions in our catalog are wrongly determined, we need to take into account a bias which may increase its probability. As stated in N06, the detection of these HII regions is not due to their own emission, but through the emission that is being blocked by them. HII regions that are blocking a column of lower synchrotron emissivity have a smaller chance of being detected in this way. This might explain why, at large distances from the observer, the only regions detected have emissivity values that appear to be enhanced. Keep in mind that this detection bias will not make the existing data at short path lengths disappear. Furthermore, accuracy of HII region distance determination is beyond the scope of this paper.


\section{Conclusions}
\label{sec:conclusions}

Summarizing our most important conclusions, we state that our observations imply either that there is a problem with the distance determination of part of the distant HII regions, or that there is a change in the emissivity contribution in either the inner region (paucity) or the outer region (enhancement) of the Galaxy. 

Assuming that the latter options are more likely, a region with a diminished emissivity contribution is likely located in the inner 8~kpc. This paucity can be caused by an outflow of CREs and/or a lack of perpendicular (to the LOS) magnetic field lines in this region. 
Conversely, a region for an emissivity enhancement has to be at least 16 kpc away from the sun in the Galactic center direction. This region has to spread out over a longitude range that is probed by our catalog (--43$^{\circ} < \ell < $ 25$^{\circ}$ ). The enhancement of synchrotron emissivity can be caused by (1) extra CREs, perhaps a single isolated source; (2) higher magnetic field strength; (3) intermittency of the magnetic field and (4) positive correlation between CRE density and the GMF, in the region behind the Galactic center.

\section*{Acknowledgements}
IP would like to acknowledge funding from the Netherlands Research School for Astronomy (NOVA). 
Carmelo Evoli, Anvar Shukurov and Elena Orlando are thanked for fruitful discussions on different parts of this work.
MH acknowledges funding from the European Research Council (ERC) under the European Union's Horizon 2020 research and innovation programme (grant agreement No 772663).
This research made extensive use of NumPy \citep{vanderWalt2011}; IPython \citep{Perez2007} and matplotlib \citep{Hunter2007}. The authors would like to thank the ADS abstract service.
The authors would like to thank the referee for suggestions that have improved this paper.
\bibliography{P2}

%
%
\longtab{5}{
\onecolumn{
\begin{longtable}{*{7}{c}}

\caption{\label{tab:catalog} The complete and updated catalog of HII regions detected in absorption. Column (1) contains the Galactic coordinates, Column (2) is the distance from the Sun to the HII region, Column (3) is the sky brightness temperature derived from the measured intensity at the observing frequency, Column (4) indicates the synchrotron brightness temperature of the column in front of the HII region (first 9 entries), and the synchrotron brightness temperature of the column behind the HII region (entries 10 and onward), Column (5) is the electron temperature of the HII region, and Column (6) is the emissivity of the column in front of the HII regions (first 9 entries) and behind the HII region (entries 10 and onward), Column (7) indicates the origin paper of the data with [1] \citet{Jones1974}, [2] \citet{Roger1999}, [3] \citet{Nord2006}, [4] \citet{Hindson2016} and [5] \citet{Su2016}.}\\
\hline
{\bf $\ell$~\textpm~b}&{\bf Distance}&{\bf T$_{\rm{obs}}$}&{\bf T$_{\rm F}$}&{\bf T$_{\rm e}$}&{\bf ${ \epsilon}_{\rm F}$}&{\bf Source}\\
&kpc&$\times10^3$~K&$\times10^3$~K&$\times10^3$~K&K~pc$^{-1}$&\\
(1)&(2)&(3)&(4)&(5)&(6)&(7)\\
\hline
\endfirsthead
\caption{continued.}\\
\hline
{\bf $\ell$~\textpm~b}&{\bf Distance}&{\bf T$_{\rm{obs}}$}&{\bf T$_{\rm B}$}&{\bf T$_{\rm e}$}&{\bf $\epsilon_{\rm B}$}&{\bf Source}\\
&kpc&$\times10^3$~K&$\times10^3$~K&$\times10^3$~K&K~pc$^{-1}$&\\
(1)&(2)&(3)&(4)&(5)&(6)&(7)\\
\hline
\endhead
\hline
\\
\endfoot
287.5--0.5&2.5\,$\pm$\,0.625&29.0\,$\pm$\,2.554&2.08\,$\pm$\,0.12&5.0\,$\pm$\,0.5\,\footnotemark[3]&0.83\,$\pm$\,0.21&[1]\\
336.7--1.3&1.3\,$\pm$\,0.325&58.0\,$\pm$\,3.605&4.59\,$\pm$\,1.48&5.0\,$\pm$\,0.5\,\footnotemark[3]&3.53\,$\pm$\,1.44&[1]\\
\hline

85.5--1.0&0.8\,$\pm$\,0.29&23.4\,\footnotemark[1]\,$\pm$\,10.0&0.66\,$\pm$\,0.2&6.0\,$\pm$\,1.0\,\footnotemark[2]&0.82\,$\pm$\,0.39&[2]\\
99.3+3.7&0.86\,$\pm$\,0.055&43.4\,\footnotemark[1]\,$\pm$\,10.0&1.41\,$\pm$\,1.58&6.0\,$\pm$\,1.0\,\footnotemark[2]&1.64\,$\pm$\,1.85&[2]\\
118.5+6.0&0.84\,\footnotemark[5]\,$\pm$\,0.084&36.1\,\footnotemark[1]\,$\pm$\,10.0&1.14\,$\pm$\,0.88&6.0\,$\pm$\,1.0\,\footnotemark[2]&1.35\,$\pm$\,1.06&[2]\\
134.8+0.9&2.2\,$\pm$\,0.097&52.0\,\footnotemark[1]\,$\pm$\,10.0&1.74\,$\pm$\,2.77&6.0\,$\pm$\,1.0\,\footnotemark[2]&0.79\,$\pm$\,1.26&[2]\\
160.1--12.3&0.4\,$\pm$\,0.039&29.9\,\footnotemark[1]\,$\pm$\,10.0&0.9\,$\pm$\,0.47&6.0\,$\pm$\,1.0\,\footnotemark[2]&2.26\,$\pm$\,1.2&[2]\\
195.1--12.0&0.4\,$\pm$\,0.04&26.4\,\footnotemark[1]\,$\pm$\,10.0&0.77\,$\pm$\,0.31&6.0\,$\pm$\,1.0\,\footnotemark[2]&1.93\,$\pm$\,0.79&[2]\\
202.9+2.2&0.8\,$\pm$\,0.25&29.1\,\footnotemark[1]\,$\pm$\,10.0&0.87\,$\pm$\,0.43&6.0\,$\pm$\,1.0\,\footnotemark[2]&1.09\,$\pm$\,0.64&[2]\\
\hline
{\bf $\ell$~\textpm~b}&{\bf Distance}&{\bf T$_{\rm{obs}}$}&{\bf T$_{\rm B}$}&{\bf T$_{\rm e}$}&{\bf $\epsilon_{\rm B}$}&{\bf Source}\\
&kpc&$\times10^3$~K&$\times10^3$~K&$\times10^3$~K&K~pc$^{-1}$&\\
(1)&(2)&(3)&(4)&(5)&(6)&(7)\\
\hline
002.3+01.4&4.7$\pm$2.35&-3.0$\pm$0.5&10.0$\pm$2.1&7.0$\pm$2.0&0.42$\pm$0.1&[3]\\
002.4+01.4&4.7$\pm$2.35&-3.6$\pm$0.5&10.6$\pm$2.1&7.0$\pm$2.0&0.45$\pm$0.1&[3]\\
003.3+00.0&16.77$\pm$8.38&-2.1$\pm$0.5&9.1$\pm$2.1&7.0$\pm$2.0&0.78$\pm$0.58&[3]\\
005.9-00.4&2.99$\pm$0.18&-6.5$\pm$0.9&13.2$\pm$1.3&6.7$\pm$1.0&0.52$\pm$0.05&[3]\\
006.0-01.2&16.1$\pm$2.56&-7.4$\pm$2.0&14.4$\pm$2.8&7.0$\pm$2.0&1.17$\pm$0.33&[3]\\
006.0-01.3&16.1$\pm$2.56&-6.4$\pm$2.0&13.4$\pm$2.8&7.0$\pm$2.0&1.09$\pm$0.32&[3]\\
006.2-01.2&0.95$\pm$0.48&-8.0$\pm$1.9&15.0$\pm$2.7&7.0$\pm$2.0&0.55$\pm$0.1&[3]\\
006.4-00.5&3.8$\pm$0.11&-3.6$\pm$1.0&10.6a$\pm$2.2&7.0$\pm$2.0&0.43$\pm$0.09&[3]\\
008.1+00.2&3.5$\pm$0.1&-8.5$\pm$1.8&15.0$\pm$2.0&6.5$\pm$1.0&0.6$\pm$0.08&[3]\\
012.7-00.2&2.4$\pm$0.16&-9.4$\pm$1.2&13.9$\pm$1.6&4.5$\pm$1.0&0.54$\pm$0.06&[3]\\
012.8-00.2&2.92$\pm$0.31&-9.3$\pm$1.2&15.3$\pm$1.6&6.0$\pm$1.0&0.61$\pm$0.06&[3]\\
012.9-00.2&2.53$\pm$0.2&-6.0$\pm$1.2&12.1$\pm$1.6&6.1$\pm$1.0&0.47$\pm$0.06&[3]\\
013.9-00.1&3.5$\pm$1.75&-4.8$\pm$1.3&11.8$\pm$2.4&7.0$\pm$2.0&0.48$\pm$0.1&[3]\\
014.0-00.1&3.6$\pm$0.06&-7.0$\pm$1.3&12.6$\pm$1.7&5.5$\pm$1.0&0.51$\pm$0.07&[3]\\
014.2-00.2&3.7$\pm$0.06&-4.4$\pm$1.4&11.4a$\pm$2.4&7.0$\pm$2.0&0.47$\pm$0.1&[3]\\
014.2-00.3&3.62$\pm$1.81&-5.8$\pm$1.4&12.8$\pm$2.4&7.0$\pm$2.0&0.52$\pm$0.11&[3]\\
014.3-00.2&3.62$\pm$1.81&-11.4$\pm$1.4&18.4$\pm$2.4&7.0$\pm$2.0&0.75$\pm$0.11&[3]\\
014.4-00.6&1.12$\pm$0.13&-4.6$\pm$1.4&11.6$\pm$2.4&7.0$\pm$2.0&0.43$\pm$0.09&[3]\\
014.5-00.6&1.98$\pm$0.99&-6.6$\pm$1.4&13.6$\pm$2.4&7.0$\pm$2.0&0.52$\pm$0.09&[3]\\
014.6+00.1&3.6$\pm$0.45&-8.1$\pm$1.4&13.4$\pm$1.7&5.3$\pm$1.0&0.55$\pm$0.07&[3]\\
015.1-00.7&2.1$\pm$0.09&-7.3$\pm$2.3&13.2$\pm$2.5&5.9$\pm$1.0&0.51$\pm$0.1&[3]\\
015.2-00.6&1.8$\pm$0.1&-8.0$\pm$2.3&17.5$\pm$2.5&9.5$\pm$1.0&0.67$\pm$0.09&[3]\\
016.9+00.8&2.7$\pm$0.07&-9.2$\pm$2.2&15.3$\pm$2.5&6.1$\pm$1.0&0.56$\pm$0.06&[3]\\
017.0+00.8&2.5$\pm$0.07&-7.0$\pm$1.6&13.1$\pm$1.9&6.1$\pm$1.0&0.51$\pm$0.08&[3]\\
017.0+00.9&2.7$\pm$0.07&-10.0$\pm$1.6&18.1$\pm$1.9&8.1$\pm$1.0&0.72$\pm$0.08&[3]\\
018.2+01.9&2.0$\pm$0.2&-4.3$\pm$1.5&10.1$\pm$1.8&5.8$\pm$1.0&0.39$\pm$0.07&[3]\\
018.3-00.3&4.0$\pm$0.05&-4.2$\pm$1.3&9.5$\pm$1.6&5.3$\pm$1.0&0.4$\pm$0.07&[3]\\
018.3+01.9&2.8$\pm$0.07&-5.6$\pm$1.5&11.4$\pm$1.8&5.8$\pm$1.0&0.45$\pm$0.07&[3]\\
018.5+01.9&2.95$\pm$1.48&-6.3$\pm$1.7&13.3$\pm$2.6&7.0$\pm$2.0&0.53$\pm$0.11&[3]\\
018.5+02.0&2.95$\pm$1.48&-4.8$\pm$1.5&11.8$\pm$2.5&7.0$\pm$2.0&0.47$\pm$0.1&[3]\\
018.6+01.9&2.7$\pm$0.5&-6.3$\pm$1.4&13.3$\pm$2.5&7.0$\pm$2.0&0.53$\pm$0.1&[3]\\
018.7+02.0&2.5$\pm$0.07&-6.0$\pm$1.5&13.0$\pm$2.5&7.0$\pm$2.0&0.51$\pm$0.1&[3]\\
018.9-00.4&4.7$\pm$0.04&-9.2$\pm$1.2&14.7$\pm$1.6&5.5$\pm$1.0&0.63$\pm$0.07&[3]\\
018.9-00.5&4.6$\pm$0.04&-3.8$\pm$1.2&9.7$\pm$1.6&5.9$\pm$1.0&0.42$\pm$0.07&[3]\\
019.0-00.0&3.8$\pm$0.07&-3.3$\pm$1.2&8.5$\pm$1.6&5.2$\pm$1.0&0.35$\pm$0.07&[3]\\
019.0-00.3&4.6$\pm$0.36&-5.1$\pm$1.2&10.3$\pm$1.6&5.2$\pm$1.0&0.44$\pm$0.07&[3]\\
019.0-00.4&3.4$\pm$1.7&-9.3$\pm$1.2&16.3$\pm$2.3&7.0$\pm$2.0&0.67$\pm$0.1&[3]\\
019.1-00.3&4.6$\pm$0.04&-6.3$\pm$1.2&10.4$\pm$1.6&4.1$\pm$1.0&0.45$\pm$0.07&[3]\\
020.2-00.9&3.65$\pm$1.82&-3.1$\pm$1.1&10.1$\pm$2.3&7.0$\pm$2.0&0.42$\pm$0.1&[3]\\
022.9-00.3&11.1$\pm$0.04&-5.2$\pm$1.1&10.3$\pm$1.5&5.1$\pm$1.0&0.63$\pm$0.09&[3]\\
023.0-00.4&4.59$\pm$1.47&-4.0$\pm$1.1&11.8$\pm$1.5&7.8$\pm$1.0&0.51$\pm$0.07&[3]\\
024.2+00.2&9.3$\pm$0.05&-4.1$\pm$1.3&11.1$\pm$2.4&7.0$\pm$2.0&0.61$\pm$0.13&[3]\\
024.4+00.1&6.2$\pm$0.05&-5.1$\pm$1.3&10.6$\pm$1.6&5.5$\pm$1.0&0.5$\pm$0.08&[3]\\
024.5+00.2&6.5$\pm$0.05&-6.1$\pm$1.3&10.8$\pm$1.7&4.7$\pm$1.0&0.52$\pm$0.08&[3]\\
024.6-00.2&5.8$\pm$0.5&-7.2$\pm$1.3&14.2$\pm$2.4&7.0$\pm$2.0&0.66$\pm$0.11&[3]\\
024.7-00.1&6.2$\pm$0.05&-6.2$\pm$1.3&13.2$\pm$2.4&7.0$\pm$2.0&0.62$\pm$0.11&[3]\\
024.7-00.2&10.3$\pm$0.04&-8.4$\pm$1.3&14.1$\pm$1.7&5.7$\pm$1.0&0.82$\pm$0.1&[3]\\
024.8+00.1&6.1$\pm$0.05&-9.8$\pm$1.4&14.7$\pm$1.7&5.0$\pm$1.0&0.69$\pm$0.08&[3]\\
025.3-00.3&11.2$\pm$0.05&-7.6$\pm$1.5&13.3$\pm$1.8&5.7$\pm$1.0&0.82$\pm$0.11&[3]\\
025.4-00.3&11.2$\pm$0.05&-8.0$\pm$1.5&16.1$\pm$1.8&8.1$\pm$1.0&1.0$\pm$0.11&[3]\\
348.7-01.0&3.38$\pm$0.3&-3.3$\pm$0.9&9.5$\pm$1.4&6.2$\pm$1.0&0.38$\pm$0.06&[3]\\
351.0+00.7&0.7$\pm$1.37&-6.8$\pm$0.9&13.8$\pm$2.2&7.0$\pm$2.0&0.5$\pm$0.08&[3]\\
351.2+00.5&1.6$\pm$1.31&-6.7$\pm$0.9&13.7$\pm$2.2&7.0$\pm$2.0&0.51$\pm$0.09&[3]\\
351.4+00.7&1.34$\pm$0.13&-6.2$\pm$0.9&12.3$\pm$1.4&6.1$\pm$1.0&0.46$\pm$0.05&[3]\\
351.5-00.5&3.3$\pm$0.1&-3.2$\pm$1.0&8.9$\pm$1.4&5.7$\pm$1.0&0.36$\pm$0.06&[3]\\
353.1+00.6&15.9$\pm$1.99&-6.0$\pm$1.1&13.0$\pm$2.3&7.0$\pm$2.0&1.04$\pm$0.25&[3]\\
353.1+00.7&15.9$\pm$1.99&-7.0$\pm$1.1&14.0$\pm$2.3&7.0$\pm$2.0&1.12$\pm$0.26&[3]\\
353.2+00.7&15.9$\pm$1.99&-11.0$\pm$1.1&18.0$\pm$2.3&7.0$\pm$2.0&1.44$\pm$0.29&[3]\\
353.2+00.9&15.9$\pm$1.99&-12.5$\pm$1.1&19.5$\pm$2.3&7.0$\pm$2.0&1.56$\pm$0.31&[3]\\
358.6-00.1&8.5$\pm$4.25&-1.8$\pm$0.4&8.8$\pm$2.0&7.0$\pm$2.0&0.44$\pm$0.14&[3]\\
359.7-00.4&1.5$\pm$0.75&-1.7$\pm$0.4&10.3$\pm$2.0&8.6$\pm$1.0&0.38$\pm$0.07&[3]\\
\hline
317.33+0.26&2.5$\pm$0.3&-0.48$\pm$0.25&11.95$\pm$3.22&7.0$\pm$2.0&0.52$\pm$0.14&[4]\\
317.62-0.40&1.9$\pm$0.6&-0.55$\pm$1.09&12.05$\pm$3.64&7.0$\pm$2.0&0.51$\pm$0.16&[4]\\
318.19-0.59&1.7$\pm$0.3&-0.43$\pm$0.18&11.86$\pm$3.21&7.0$\pm$2.0&0.5$\pm$0.13&[4]\\
321.15-0.55&3.8$\pm$0.5&-0.83$\pm$1.25&8.51$\pm$2.55&4.5$\pm$1.0&0.39$\pm$0.12&[4]\\
322.19+0.57&1.8$\pm$0.9&-0.39$\pm$0.39&11.79$\pm$3.25&7.0$\pm$2.0&0.49$\pm$0.14&[4]\\
326.23+0.72&3.0$\pm$0.4&-0.47$\pm$0.18&8.74$\pm$1.62&5.0$\pm$1.0&0.37$\pm$0.07&[4]\\
326.67+0.57&2.4$\pm$0.3&-1.68$\pm$0.62&13.85$\pm$3.34&7.0$\pm$2.0&0.57$\pm$0.14&[4]\\
327.18-0.60&2.9$\pm$1.0&-1.39$\pm$0.36&13.39$\pm$3.24&7.0$\pm$2.0&0.56$\pm$0.14&[4]\\
329.36+0.12&7.3$\pm$0.1&-0.79$\pm$1.57&12.91$\pm$2.98&7.3$\pm$1.0&0.66$\pm$0.15&[4]\\
330.89-0.37&3.7$\pm$0.4&-0.16$\pm$0.16&8.07$\pm$1.62&4.9$\pm$1.0&0.35$\pm$0.07&[4]\\
331.15-0.52&4.3$\pm$0.4&-1.03$\pm$0.45&8.82$\pm$1.75&4.5$\pm$1.0&0.39$\pm$0.08&[4]\\
332.98+1.78&2.1$\pm$1.4&-0.71$\pm$0.39&12.3$\pm$3.25&7.0$\pm$2.0&0.49$\pm$0.13&[4]\\
333.01-0.62&3.1$\pm$1.1&-5.07$\pm$1.59&19.27$\pm$4.07&7.0$\pm$2.0&0.8$\pm$0.17&[4]\\
333.04+2.03&1.6$\pm$0.6&-2.01$\pm$0.8&12.94$\pm$2.05&6.1$\pm$1.0&0.51$\pm$0.08&[4]\\
333.06-0.45&3.0$\pm$0.3&-0.91$\pm$0.35&12.63$\pm$3.24&7.0$\pm$2.0&0.52$\pm$0.13&[4]\\
333.07+0.02&2.5$\pm$1.1&-5.51$\pm$2.36&19.97$\pm$4.94&7.0$\pm$2.0&0.81$\pm$0.2&[4]\\
333.20-0.10&2.4$\pm$0.3&-1.49$\pm$0.6&13.55$\pm$3.33&7.0$\pm$2.0&0.55$\pm$0.13&[4]\\
333.29-0.30&3.3$\pm$1.1&-11.19$\pm$2.8&29.03$\pm$5.49&7.0$\pm$2.0&1.21$\pm$0.24&[4]\\
333.33-0.39&2.7$\pm$0.3&-2.7$\pm$0.86&15.48$\pm$3.48&7.0$\pm$2.0&0.63$\pm$0.14&[4]\\
333.61-0.09&3.0$\pm$0.5&-2.31$\pm$1.03&14.87$\pm$3.59&7.0$\pm$2.0&0.61$\pm$0.15&[4]\\
333.64-0.22&3.2$\pm$0.4&-3.06$\pm$0.81&14.79$\pm$2.05&6.2$\pm$1.0&0.61$\pm$0.09&[4]\\
333.71-0.46&11.8$\pm$0.4&-3.08$\pm$2.05&8.91$\pm$3.65&2.5$\pm$1.0&0.58$\pm$0.24&[4]\\
335.78+0.01&2.2$\pm$0.3&-1.79$\pm$0.81&14.03$\pm$3.45&7.0$\pm$2.0&0.56$\pm$0.14&[4]\\
336.52-1.50&1.8$\pm$0.3&-0.57$\pm$0.29&12.09$\pm$3.23&7.0$\pm$2.0&0.47$\pm$0.13&[4]\\
336.59-1.81&2.5$\pm$0.7&-0.04$\pm$0.19&11.24$\pm$3.21&7.0$\pm$2.0&0.45$\pm$0.13&[4]\\
338.95+0.59&2.1$\pm$0.3&-1.22$\pm$0.53&13.12$\pm$3.3&7.0$\pm$2.0&0.51$\pm$0.13&[4]\\
339.18-0.42&3.0$\pm$0.4&-14.98$\pm$3.57&32.86$\pm$5.91&5.6$\pm$1.0&1.33$\pm$0.24&[4]\\
\hline
317.988--00.754&3.6\,$\pm$\,1.1&--1.33\,$\pm$\,0.28&9.47\,$\pm$\,0.74&4.6\,$\pm$\,0.37&0.43\,$\pm$\,0.04&[5]\\
322.036+00.625&3.5\,$\pm$\,3.5&--0.52\,$\pm$\,0.23&12.47\,$\pm$\,0.64&7.29\,$\pm$\,0.33&0.55\,$\pm$\,0.09&[5]\\
322.220+00.504&3.5\,$\pm$\,3.5&--0.38\,$\pm$\,0.28&12.25\,$\pm$\,0.69&7.29\,$\pm$\,0.33&0.54\,$\pm$\,0.09&[5]\\
326.270+00.783&3.0\,$\pm$\,0.4&--2.56\,$\pm$\,0.58&15.79\,$\pm$\,1.07&7.33\,$\pm$\,0.33&0.67\,$\pm$\,0.05&[5]\\
326.643+00.514&3.0\,$\pm$\,0.4&--1.35\,$\pm$\,0.58&13.84\,$\pm$\,1.07&7.32\,$\pm$\,0.33&0.59\,$\pm$\,0.05&[5]\\
327.300--00.548&3.2\,$\pm$\,0.4&--1.99\,$\pm$\,0.42&12.92\,$\pm$\,0.89&6.1\,$\pm$\,0.36&0.55\,$\pm$\,0.04&[5]\\
327.991--00.087&3.6\,$\pm$\,1.8&--0.79\,$\pm$\,0.35&10.84\,$\pm$\,0.81&6.0\,$\pm$\,0.36&0.47\,$\pm$\,0.05&[5]\\
328.572--00.527&3.4\,$\pm$\,0.4&--2.38\,$\pm$\,0.38&15.28\,$\pm$\,0.81&7.19\,$\pm$\,0.33&0.65\,$\pm$\,0.04&[5]\\
331.365+00.521&11.8\,$\pm$\,5.9&--3.3\,$\pm$\,0.52&12.93\,$\pm$\,1.0&4.8\,$\pm$\,0.34&0.85\,$\pm$\,0.33&[5]\\
332.145--00.452&3.7\,$\pm$\,0.4&--1.64\,$\pm$\,0.28&13.87\,$\pm$\,0.68&7.05\,$\pm$\,0.32&0.59\,$\pm$\,0.03&[5]\\
332.657--00.622&3.3\,$\pm$\,0.4&--2.1\,$\pm$\,0.78&14.77\,$\pm$\,1.34&7.15\,$\pm$\,0.32&0.62\,$\pm$\,0.06&[5]\\
332.762--00.595&3.8\,$\pm$\,0.4&--2.12\,$\pm$\,0.78&14.58\,$\pm$\,1.34&7.01\,$\pm$\,0.32&0.62\,$\pm$\,0.06&[5]\\
332.978+00.773&3.8\,$\pm$\,0.5&--2.25\,$\pm$\,0.33&9.98\,$\pm$\,0.76&4.0\,$\pm$\,0.35&0.43\,$\pm$\,0.03&[5]\\
333.011--00.441&3.6\,$\pm$\,0.4&--1.89\,$\pm$\,0.47&14.29\,$\pm$\,0.9&7.06\,$\pm$\,0.32&0.61\,$\pm$\,0.04&[5]\\
333.093+01.966&1.6\,$\pm$\,0.6&--1.13\,$\pm$\,0.48&14.05\,$\pm$\,0.95&7.67\,$\pm$\,0.35&0.55\,$\pm$\,0.04&[5]\\
333.627--00.199&3.2\,$\pm$\,0.4&--3.51\,$\pm$\,0.54&17.03\,$\pm$\,1.0&7.16\,$\pm$\,0.32&0.71\,$\pm$\,0.04&[5]\\
337.957--00.474&3.1\,$\pm$\,1.6&--1.13\,$\pm$\,0.38&10.74\,$\pm$\,0.83&5.6\,$\pm$\,0.35&0.44\,$\pm$\,0.04&[5]\\
338.706+00.645&4.3\,$\pm$\,0.4&--2.5\,$\pm$\,0.59&14.78\,$\pm$\,1.07&6.76\,$\pm$\,0.31&0.63\,$\pm$\,0.05&[5]\\
338.911+00.615&4.4\,$\pm$\,0.4&--2.67\,$\pm$\,0.59&15.01\,$\pm$\,1.07&6.73\,$\pm$\,0.31&0.64\,$\pm$\,0.05&[5]\\
338.934--00.067&3.2\,$\pm$\,0.4&--2.36\,$\pm$\,0.42&15.1\,$\pm$\,0.85&7.1\,$\pm$\,0.32&0.62\,$\pm$\,0.04&[5]\\
339.109--00.233&6.5\,$\pm$\,3.3&--2.05\,$\pm$\,0.58&9.98\,$\pm$\,1.06&4.2\,$\pm$\,0.32&0.47\,$\pm$\,0.09&[5]\\
339.134--00.377&3.0\,$\pm$\,0.4&--3.47\,$\pm$\,0.58&16.97\,$\pm$\,1.06&7.16\,$\pm$\,0.32&0.69\,$\pm$\,0.04&[5]\\
340.216+00.424&4.4\,$\pm$\,2.2&--2.77\,$\pm$\,0.59&12.09\,$\pm$\,1.08&4.8\,$\pm$\,0.33&0.52\,$\pm$\,0.07&[5]\\
340.678--01.049&2.3\,$\pm$\,2.3&--2.25\,$\pm$\,0.57&15.37\,$\pm$\,1.05&7.38\,$\pm$\,0.33&0.6\,$\pm$\,0.07&[5]\\
340.780--01.022&2.3\,$\pm$\,0.6&--2.6\,$\pm$\,0.57&15.93\,$\pm$\,1.05&7.38\,$\pm$\,0.33&0.62\,$\pm$\,0.04&[5]\\
340.862--00.870&2.3\,$\pm$\,2.3&--1.47\,$\pm$\,0.47&14.13\,$\pm$\,0.91&7.38\,$\pm$\,0.33&0.55\,$\pm$\,0.06&[5]\\
341.090--00.017&3.2\,$\pm$\,3.2&--2.75\,$\pm$\,0.3&15.68\,$\pm$\,0.7&7.07\,$\pm$\,0.32&0.64\,$\pm$\,0.09&[5]\\
342.277+00.311&9.6\,$\pm$\,4.8&--3.22\,$\pm$\,0.66&11.37\,$\pm$\,1.18&3.9\,$\pm$\,0.32&0.62\,$\pm$\,0.17&[5]\\
343.480--00.043&13.4\,$\pm$\,7.4&--2.22\,$\pm$\,0.42&16.48\,$\pm$\,0.88&8.1\,$\pm$\,0.35&1.13\,$\pm$\,0.57&[5]\\
343.914--00.646&2.8\,$\pm$\,1.4&--0.94\,$\pm$\,0.33&13.0\,$\pm$\,0.76&7.2\,$\pm$\,0.35&0.52\,$\pm$\,0.04&[5]\\
345.094--00.779&2.1\,$\pm$\,2.1&--3.59\,$\pm$\,0.62&17.59\,$\pm$\,1.13&7.43\,$\pm$\,0.33&0.68\,$\pm$\,0.07&[5]\\
345.202+01.027&1.1\,$\pm$\,0.6&--4.13\,$\pm$\,1.22&14.26\,$\pm$\,1.95&4.8\,$\pm$\,0.12&0.53\,$\pm$\,0.07&[5]\\
345.235+01.408&8.0\,$\pm$\,4.0&--1.94\,$\pm$\,0.68&12.68\,$\pm$\,1.22&6.0\,$\pm$\,0.35&0.63\,$\pm$\,0.14&[5]\\
345.410--00.953&2.6\,$\pm$\,0.6&--2.26\,$\pm$\,0.81&14.72\,$\pm$\,1.29&6.96\,$\pm$\,0.05&0.58\,$\pm$\,0.05&[5]\\
348.261+00.485&1.8\,$\pm$\,1.8&--3.74\,$\pm$\,0.64&17.99\,$\pm$\,1.15&7.53\,$\pm$\,0.34&0.68\,$\pm$\,0.06&[5]\\
348.691--00.826&3.4\,$\pm$\,0.3&--1.24\,$\pm$\,0.66&9.64\,$\pm$\,1.92&4.8\,$\pm$\,1.0&0.39\,$\pm$\,0.08&[5]\\
348.710--01.044&3.4\,$\pm$\,0.3&--1.97\,$\pm$\,0.57&13.04\,$\pm$\,1.83&6.2\,$\pm$\,1.0&0.52\,$\pm$\,0.07&[5]\\
350.991--00.532&13.7\,$\pm$\,6.9&--1.62\,$\pm$\,0.62&12.33\,$\pm$\,1.14&6.1\,$\pm$\,0.35&0.84\,$\pm$\,0.4&[5]\\
350.995+00.654&0.6\,$\pm$\,0.3&--6.76\,$\pm$\,1.34&27.67\,$\pm$\,2.21&10.57\,$\pm$\,0.34&1.0\,$\pm$\,0.08&[5]\\
351.130+00.449&1.4\,$\pm$\,0.7&--7.62\,$\pm$\,1.64&22.78\,$\pm$\,2.62&6.65\,$\pm$\,0.07&0.85\,$\pm$\,0.1&[5]\\
351.311+00.663&1.3\,$\pm$\,0.1&--6.84\,$\pm$\,0.69&23.23\,$\pm$\,1.24&7.71\,$\pm$\,0.35&0.86\,$\pm$\,0.05&[5]\\
351.383+00.737&1.3\,$\pm$\,0.1&--7.43\,$\pm$\,0.68&27.35\,$\pm$\,1.09&9.7\,$\pm$\,0.09&1.01\,$\pm$\,0.04&[5]\\
351.516--00.540&3.3\,$\pm$\,3.3&--3.9\,$\pm$\,0.68&15.33\,$\pm$\,1.93&5.7\,$\pm$\,1.0&0.61\,$\pm$\,0.11&[5]\\
351.688--01.169&14.2\,$\pm$\,1.0&--3.05\,$\pm$\,0.49&15.23\,$\pm$\,0.86&6.49\,$\pm$\,0.21&1.07\,$\pm$\,0.1&[5]\\
353.038+00.581&1.1\,$\pm$\,1.1&--5.39\,$\pm$\,1.12&21.03\,$\pm$\,1.87&7.78\,$\pm$\,0.35&0.77\,$\pm$\,0.08&[5]\\
353.076+00.287&0.7\,$\pm$\,1.5&--6.72\,$\pm$\,1.5&19.33\,$\pm$\,2.4&5.39\,$\pm$\,0.1&0.7\,$\pm$\,0.09&[5]\\
353.092+00.857&1.0\,$\pm$\,2.0&--5.72\,$\pm$\,1.12&20.47\,$\pm$\,1.89&7.1\,$\pm$\,0.4&0.75\,$\pm$\,0.09&[5]\\

\end{longtable}

\footnotetext[1]{Observed values not given in original paper; values calculated in P19.}
\footnotetext[2]{Uncertainty absent in paper, we adopt 1.0\,$\times\,10^3$~K.}
\footnotetext[3]{Electron temperatures updated with values from \citet{Azcarate1990}.}

}}

\end{document}